\documentclass[prd,superscriptaddress,twocolumn,nofootinbib,showpacs,preprintnumbers,floatfix]{revtex4}

\usepackage{mathrsfs}
\usepackage{amsfonts,amssymb,amsmath}
\usepackage{graphicx,epsfig}
\usepackage{multirow}
\usepackage{verbatim}

\def\fsu5{$\cal{F}$-$SU(5)$}
\def\bfsu5{$\boldsymbol{\mathcal{F}}$-$\boldsymbol{SU(5)}$}

\def\m1half{$M_{1/2}$}
\def\m3half{$M_{3/2}$}
\def\m32{$M_{32}$}

\def\fb{${\rm fb}^{-1}$~}

\def\mt2{$M_{T2}$}
\def\x2{$\chi^2$}
\def\2b{$M_{T2}b$}
\def\sb{$S/\sqrt{B+1}$~}
\def\bs0{$B_S^0 \rightarrow \mu^+ \mu^-$}

\newcommand{\beqa}{\begin{eqnarray}} 
\newcommand{\eeqa}{\end{eqnarray}}
\newcommand{\Bbsmm}{\ensuremath{{\mathcal Br}(\bsmm)}}
\newcommand{\bsmm}{\ensuremath{B^0_s\to\mu^+\mu^-}}

\begin{document}

\title{Correlating LHCb $B_s^0 \rightarrow \mu^+ \mu^-$ Results with\\the ATLAS-CMS Multijet Supersymmetry Search}

\author{Tianjun Li}

\affiliation{State Key Laboratory of Theoretical Physics and Kavli Institute for Theoretical Physics China (KITPC),
Institute of Theoretical Physics, Chinese Academy of Sciences, Beijing 100190, P. R. China}

\affiliation{George P. and Cynthia W. Mitchell Institute for Fundamental Physics and Astronomy, Texas A$\&$M University, College Station, TX 77843, USA}

\author{James A. Maxin}

\affiliation{George P. and Cynthia W. Mitchell Institute for Fundamental Physics and Astronomy, Texas A$\&$M University, College Station, TX 77843, USA}

\author{Dimitri V. Nanopoulos}

\affiliation{George P. and Cynthia W. Mitchell Institute for Fundamental Physics and Astronomy, Texas A$\&$M University, College Station, TX 77843, USA}

\affiliation{Astroparticle Physics Group, Houston Advanced Research Center (HARC), Mitchell Campus, Woodlands, TX 77381, USA}

\affiliation{Academy of Athens, Division of Natural Sciences, 28 Panepistimiou Avenue, Athens 10679, Greece}

\author{Joel W. Walker}

\affiliation{Department of Physics, Sam Houston State University, Huntsville, TX 77341, USA}


\begin{abstract}

We show that the No-Scale Flipped $SU$(5) construction is transparently consistent with recent LHCb
results for \bs0 decays, due primarily to suppression from the rather small value of $\tan \beta \sim 20$
that is globally enforced across the model space.  This fact should be interpreted in conjunction
with the demonstrated evasion of mass limits from the ATLAS and CMS SUSY searches and the more important
potential explanation of small observed excesses in the multijet data.  The No-Scale
Flipped $SU$(5) benchmark model that best fits these excesses has a gaugino mass scale of
$M_{1/2} = 708$~GeV, which drives masses for the bino-dominated LSP $m_{\widetilde{\chi}_1^0} = 143.4$~GeV,
light stop $m_{\widetilde{t}_1} = 786$~GeV, gluino $m_{\widetilde{g}} = 952$~GeV, and heavy squark
$m_{\widetilde{u}_L} = 1490$ GeV.  The corresponding total prediction for the rare B-decay of
$Br$(\bs0) = 3.5 $\times 10^{-9}$ suggests that the SUSY contribution may indeed be much
smaller than that expected from the Standard Model in this framework, fitting quite comfortably within the
very tightly constrained region remaining viable after the most recent LHCb measurements.

\end{abstract}


\pacs{11.10.Kk, 11.25.Mj, 11.25.-w, 12.60.Jv}

\preprint{ACT-09-12, MIFPA-12-23}

\maketitle


Upon completion of several ATLAS and CMS 5 \fb supersymmetry (SUSY) searches of the
total 2011 LHC data harvest~\cite{ATLAS-CONF-2012-033,ATLAS-CONF-2012-037,ATLAS-CONF-2012-041,CMS-PAS-SUS-11-020,Chatrchyan:2012qka,SUS-12-002},
interesting correlations~\cite{Li:2012tr,Li:2012ix} began to emerge between event signatures
registered by the collaborations and a high-energy model framework known as \fsu5 (See
Refs.~\cite{Maxin:2011hy,Li:2011xu,Li:2011rp,Li:2011fu,Li:2011av,Li:2011ab,Li:2012hm,Li:2012tr,Li:2012ix,Li:2012qv,Li:2012jf}
and all references therein), which combines the No-Scale Flipped $SU$(5) grand unified theory (GUT) with
extra vector-like particles (flippons).  Despite an absence of any conclusive signs of supersymmetry thus far at
the LHC, we presented the case~\cite{Li:2012tr,Li:2012ix} that those studies with curious event excesses
over the expected Standard Model (SM) background all implicate a consistent narrow swath of the
\fsu5 SUSY mass scale, subsequent to an embedding of the collaboration selection strategies into that construction.
The largest significance in signal strength was observed in the ATLAS multijet
realm~\cite{ATLAS-CONF-2012-033,ATLAS-CONF-2012-037}, permitting determination of prospective
best fit SUSY masses for a bino-dominated LSP $m_{\widetilde{\chi}_1^0} = 143.4$~GeV, the light stop
$m_{\widetilde{t}_1} = 786$~GeV, gluino $m_{\widetilde{g}} = 952$~GeV, and heavy squark
$m_{\widetilde{u}_L} = 1490$~GeV~\cite{Li:2012tr}. Intriguingly, the most meaningful signal strength within data
reported by CMS also prevailed in the multijet domain~\cite{SUS-12-002}. Given that the \fsu5
supersymmetric event landscape at the LHC is anticipated to be dominated by
multijets~\cite{Maxin:2011hy,Li:2011rp,Li:2011fu,Li:2011av,Li:2011ab,Li:2012hm}, the
cumulative fidelity of the proffered explanation would be enhanced by the systematic preference of
a budding phase of SUSY event production for the multijet search space.

These mounting correlations compel the undertaking of fresh consistency checks against recently
improved B-decay constraints, specifically those from the flavour changing neutral current process \bs0~\cite{Huang:1998vb},
where the initial quark content is $(s,\overline{b})$. We take the branching ratio from Refs.~\cite{Arnowitt:2002cq,Beskidt:2011qf}, which we write in the form

\beqa  && \Bbsmm  =   {{2\tau_B m_B^5}\over{64\pi}}f^2_{B_s}\sqrt{1-{{4m_{\mu}^2}\over{m_B^2}}}
	\nonumber \\
	& &
 \left[ \left( 1-{{4m_{\mu}^2}\over{m_B^2}} \right)
	\left| {{(C_S-C_S')}\over{(m_b+m_s)}} \right|^2+
 \left| {{(C_P-C_P')}\over{(m_b+m_s)}}+ 	\right. \right. \nonumber\\
	& & \left. \left.
	2{m_{\mu}\over m_{B_s}^2}(C_A-C_A') \right|^2
	\right]~~
\label{eq:bsmm_br_msugra}
\eeqa
where $f_{B_s}$  is the $B_s$ decay constant, $m_{B}$ is the $B$ meson mass, and $\tau_B$ is the mean life. 
The factors $C_A$, $C^{\prime}_A$ are primarily determined by the Standard Model diagrams, whereas $C_S$, $C^{\prime}_S$, $C_P$,
$C^{\prime}_P$ include the SUSY loop contributions resulting from diagrams relating to particles such as, for example, the light stop, chargino, sneutrino, and Higgs bosons.

The LHCb has undertaken measurements of rare B-decay
processes with unprecedented precision, establishing an upper bound on the branching ratio of the \bs0
process of $Br$(\bs0)~$< 4.5(3.8) \times 10^{-9}$ at the 95\% (90\%) confidence
level~\cite{Aaij:2012ac}. With well-defined predictions in the Standard Model of
$Br$(\bs0)~$ = (3.2 \pm 0.2) \times 10^{-9}$~\cite{Buras:2010mh,Buras:2010wr}, where the loop-level process
employs a virtual $W$-boson to transmute the quark content and facilitate a $t\bar{t} \to Z^0$ fusion
event, severe constraints may potentially be placed upon the viable parameter spaces of any candidate
SUSY framework~\cite{Straub:2012jb}.  In particular, the allowed SUSY contribution to
these rare B-decays is in a state of rapid contraction following recent
CMS~\cite{Chatrchyan:2012rg} and LHCb~\cite{Aaij:2012ac} high-precision measurements.
Simultaneously surviving the swiftly escalating ATLAS and CMS SUSY mass spectrum
constraints while making only compulsorily insubstantial SUSY contributions to \bs0
is truly becoming a fine needle to thread.

The vector-like flippon multiplets will contribute to the $B$ rare decays to $\mu^+ \mu^-$ since they will
contribute to the two axial-current operators $O_{10}$ and $O'_{10}$ from the mixings with the SM
fermions~\cite{Li:2012xz}. Thus, the process $B^0_s \to \mu^+ \mu^-$ will give strong constraints on
the tree-level flavour changing neutral current effects~\cite{Li:2012xz}. Note that without flippon
contributions, we have $Br(B^0_s \to \mu^+ \mu^-) = 3.5 \times 10^{-9}$ for the benchmark points of
Ref.~\cite{Li:2012tr,Li:2012qv,Li:2012jf}. Thus, we must suppress the flippon contributions by some combined effect of 
i) the natural heaviness of the multiplets, around a few times the TeV scale, and ii) an assumption
that the mixings between the flippons and the SM fermions are relatively small.

\begin{figure*}[htp]
        \centering
        \includegraphics[width=0.85\textwidth]{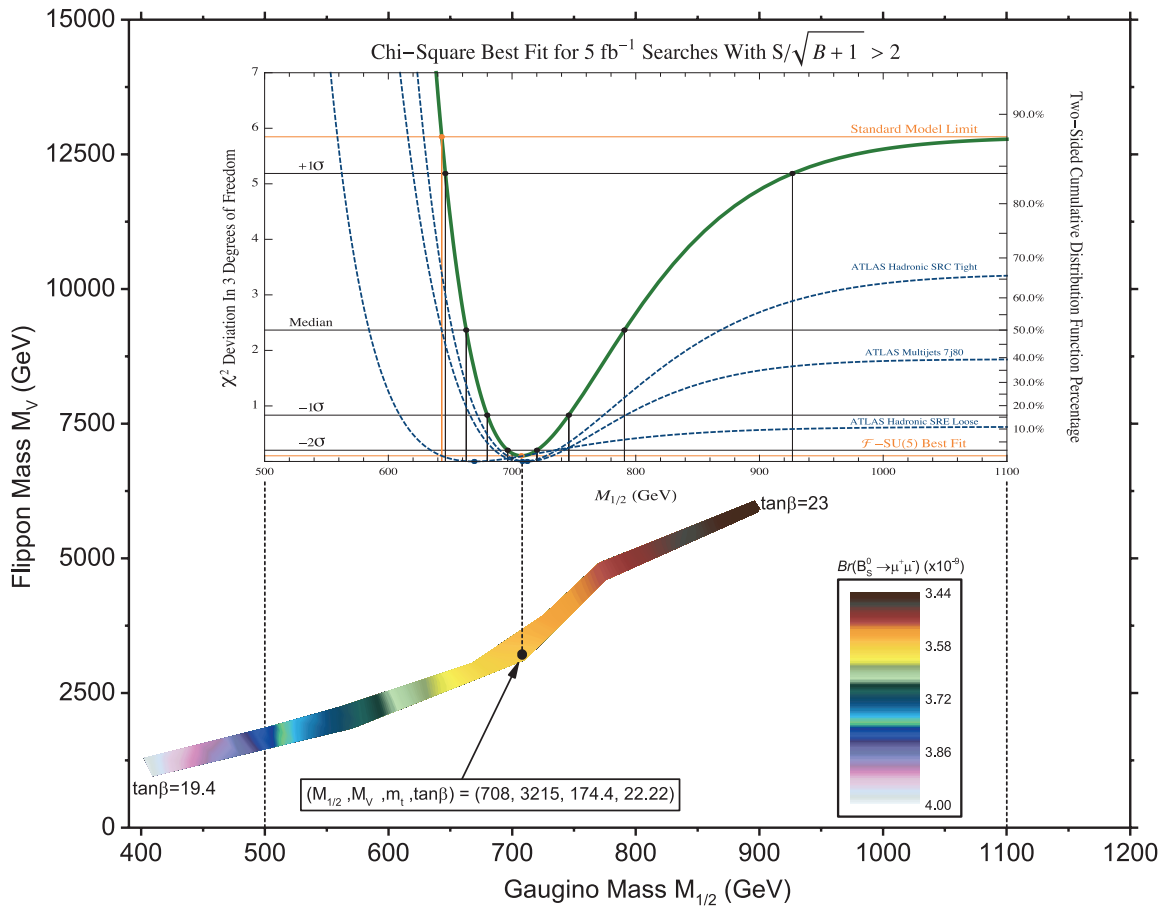}
        \caption{We depict the experimentally viable parameter space of No-Scale \fsu5 as a function of the gaugino mass $M_{1/2}$
and flippon mass $M_V$. The surviving model space after application of the bare-minimal constraints of Ref.~\cite{Li:2011xu}
and Higgs boson mass calculations of Ref.~\cite{Li:2011ab} is illustrated by the narrow strip with the smoothly contoured
color gradient. The gradient represents the total branching ratio (SM+SUSY) of the B-decay process \bs0, numerically scaled
as displayed in the color bar legend. The inset diagram (with linked horizontal scale) is the multi-axis cumulative \x2
fitting of Ref.~\cite{Li:2012tr}, depicting the best SUSY mass fit and Standard Model limit of only those ATLAS and CMS
SUSY searches exhibiting a signal significance of \sb $>$ 2. Shown is the best fit benchmark of Ref.~\cite{Li:2012tr}
at $M_{1/2}$ = 708 GeV, with $Br$(\bs0) = $3.5 \times 10 ^{-9}$, consistent with the recent LHCb measurements
of $Br$(\bs0) $<$ 4.5 (3.8) $\times 10^{-9}$ at 95\% (90\%) confidence level.}
        \label{fig:wedge}
\end{figure*}

The \fsu5 experimentally viable model space perimeter is shaped by application of the bare-minimal
experimental constraints defined in Ref.~\cite{Li:2011xu}, which is defined by simultaneous
consistency with i) the dynamically established high-scale boundary conditions $M_0$=A=$B_{\mu}$=0
of No-Scale Supergravity, ii) radiative electroweak symmetry breaking, iii) precision LEP constraints
on the lightest CP-even Higgs boson $m_{h}$~\cite{Barate:2003sz,Yao:2006px} and other light SUSY
chargino and neutralino mass content, iv) a top-quark mass $172.2~{\rm GeV} \leq m_{\rm t} \leq
174.4~{\rm GeV}$, and v) a single, neutral supersymmetric cold dark-matter (CDM) candidate providing a
relic density within the 7-year WMAP limits $0.1088 \leq \Omega_{\rm CDM} \leq
0.1158$~\cite{Komatsu:2010fb}. These constraints represent those experiments that are regarded as
exhibiting the greatest stability, with conclusions possessing the broadest acceptance. Moreover, we
superpose on the model space derived out of the bare-minimal constraints upper and lower boundaries on the
light Higgs boson mass of 124 $\le m_h \le$ 126 GeV, reflecting the recent 5$\sigma$ over background discovery observed by ATLAS~\cite{:2012gk}, CMS~\cite{:2012gu}, and CDF/D\O~\cite{Aaltonen:2012qt}. The flippons fill an essential role
in this context by coupling to the Higgs boson and naturally generating a 3--4 GeV upward shift~\cite{Li:2011ab} to
$m_h$, facilitating a physical Higgs boson mass in excellent conformity with the observations.

The \fsu5 model space remaining after implementation of the bare-minimal constraints plus the
124--126 GeV Higgs mass limits is illustrated in Figure (\ref{fig:wedge}) as a function of the gaugino
mass $M_{1/2}$ and flippon mass $M_V$, encompassing a narrow sliver confined to the region of 400 $\le
M_{1/2} \le$ 900 GeV, 19.4 $\le$ tan$\beta$ $\le$ 23, and 950 $\le M_V \le$ 6000 GeV. The lowermost boundary
at $M_{1/2}$ = 400 GeV is demanded by the LEP constraints, while the uppermost boundary at $M_{1/2}$ = 900
GeV is a consequence of a charged stau LSP exclusion around tan$\beta \simeq$ 23. The SUSY particle masses
and B-decay branching ratios are calculated with {\tt MicrOMEGAs 2.1}~\cite{Belanger:2008sj},
applying a proprietary modification of the {\tt SuSpect 2.34}~\cite{Djouadi:2002ze} codebase to run
the flippon-enhanced renormalization group equations (RGEs).

We inset into Figure (\ref{fig:wedge}) the multi-axis cumulative \x2 fitting of
Ref.~\cite{Li:2012tr}, linked to the horizontal axis $M_{1/2}$. Clearly illustrated is the well of the
\x2 at $M_{1/2}$ = 708 GeV, representing the best fit SUSY mass to those ATLAS SUSY searches demonstrating a
signal significance of \sb $>$ 2. The smoothly graded contours of color depict the value of $Br$(\bs0). The
rate for the SUSY contribution to \bs0 is proportional to the sixth power of tan$\beta$, and due to the fact
that \fsu5 globally enforces a relatively small value of 19.4 $\le$ tan$\beta$ $\le$ 23, the \fsu5
model space within the median fit of the \x2 well resides at $Br$(\bs0) $\le 3.6 \times 10^{-9}$, with
$Br$(\bs0) = 3.5 $\times 10^{-9}$ at $M_{1/2}$ = 708 GeV, both comfortably in accordance with the very
tight LHCb constraint of $Br$(\bs0) $<$ 4.5 (3.8) $\times 10^{-9}$ at the 95\% (90\%) confidence level. So
indeed, it seems No-Scale \fsu5 has successfully threaded the needle of ATLAS and CMS SUSY mass constraints
in parallel with an exceptionally small SUSY contribution to \bs0, the combination deemed to be so
intractable for the typical SUSY framework.

\textbf{Conclusions}--The LHC has amassed a total of 5 \fb of integrated luminosity at $\sqrt{s} = 7$~TeV through
the close of 2011. Consequently, the entire landscape of supersymmetric models has dwindled considerably, as the
increasing SUSY mass limits have invalidated many prior contenders. For those few high-energy frameworks left standing, the
unprecedented precision of the LHCb measurements of the B-decay process \bs0 could have been the final blow. In spite
of this dim outlook, we showed here that No-Scale \fsu5, which has already demonstrated the capacity to evade the encroaching
ATLAS and CMS SUSY mass constraints while perhaps moreover {\it explaining} the origin of small excesses in the 5 \fb multijet data observations,
is in fact further bolstered by the new LHCb results. Due to the relatively small globally allowed range of
$19.4 \le \tan\beta \le 23$, the \fsu5 SUSY contribution to $Br$(\bs0), which is proportional to the sixth power of tan$\beta$,
is much smaller than the effect expected within the Standard Model. Thus, a large value of $Br$(\bs0) could have dealt a
very damaging hit to \fsu5. On the contrary, the now apparent insubstantial SUSY contribution measured at the LHCb is very
consistent with that required in a No-Scale \fsu5 universe. Indeed, we demonstrated that the SUSY mass spectrum of an
LSP $m_{\widetilde{\chi}_1^0} = 143.4$~GeV, light stop $m_{\widetilde{t}_1} = 786$~GeV, gluino $m_{\widetilde{g}} = 952$~GeV,
and heavy squark $m_{\widetilde{u}_L}$ = 1490 GeV, which can efficiently explain the ATLAS multijet data observation excesses,
exhibits a B-decay of $Br$(\bs0) = 3.5 $\times 10^{-9}$, well within the quite constrained LHCb result of
$Br$(\bs0) $<$ 4.5 (3.8) $\times 10^{-9}$ at the 95\% (90\%) confidence level. Whether nature is herself truly described by
No-Scale \fsu5 remains, for now, beyond our capacity to establish; however, it is becoming increasingly clear
that her actions, spanning a broad and non-trivially correlated space of observations, conform remarkably well
to the predictions that this model makes.  This elegant evasion of myriad potential pitfalls, the rare B-decay limits
being but one example among many, serves to highlight the sharp differences in phenomenology
that exist between the \fsu5 framework and the traditional CMSSM/mSUGRA constructions.


\begin{acknowledgments}
This research was supported in part
by the DOE grant DE-FG03-95-Er-40917 (TL and DVN),
by the Natural Science Foundation of China
under grant numbers 10821504, 11075194, and 11135003 (TL),
by the Mitchell-Heep Chair in High Energy Physics (JAM),
and by the Sam Houston State University
2011 Enhancement Research Grant program (JWW).
We also thank Sam Houston State University
for providing high performance computing resources.
\end{acknowledgments}


\bibliography{bibliography}

\end{document}